# Electromagnetic signal propagation through lossy media via surface electromagnetic waves

Igor I. Smolyaninov *, Quirino Balzano, Vera N. Smolyaninova, Daryna Soloviova

**Abstract:** A theory of surface electromagnetic waves in gradient media exhibiting arbitrary surface gradients of dielectric permittivity and magnetic permeability has been developed. Novel low-loss propagating surface wave solutions have been found in the gradient media in which both dielectric permittivity and magnetic permeability are dominated by their imaginary parts. Several examples of gradient geometries in which the surface wave problem may be solved analytically have been found. Examples of practically useful surface wave geometries spanning from radio communication underwater to UV nanophotonics have been demonstrated.

**Keywords:** surface electromagnetic wave; gradient media; underwater communication; UV nanophotonics; PT-symmetry.

## 1 Introduction

Surface electromagnetic wave (SEW) solutions of macroscopic Maxwell equations, such as surface plasmon polaritons (SPP) at sharp metal-dielectric boundaries [1] and Zenneck waves at sharp boundaries between high-loss and low-loss dielectrics [2] have been known for quite a while. More recently, SEW solutions were also discovered in the low-loss [3] and even in very high-loss [4,5] gradient dielectric media, such as the surface of seawater and various underground sediment layers. The latter discovery was extremely surprising and practically useful since such SEWs may be used in broadband underwater radio [6] and video [7] communication. Another potential application of this result is UV nanophotonics [8], since virtually all materials behave as very high loss dielectrics at UV frequencies. In addition, this surprising result also has broad fundamental physics implications, since it complements recent observations of loss-enhanced transmission due to PT-symmetry in non-Hermitian optical systems [9].

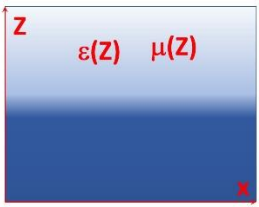

**Fig. 1:** Gradient interface problem considered in this paper. The dielectric permittivity ε(z) and the magnetic permeability μ(z) of the medium are continuous and depend only on z coordinate, which is illustrated schematically by halftones.

Unfortunately, these newly found SEW solutions still suffer from relatively short propagation range. Even though their propagation range is considerably longer than their wavelength, it remains short in practical terms, which considerably limits their applications. The goal of this paper is to extend the theoretical consideration in refs.[4,5] to a more general case in which a highly lossy gradient medium exhibits surface gradients of both its dielectric permittivity ε(z) and its magnetic permeability μ(z), as illustrated in Fig.1. It will be demonstrated that novel lower-loss propagating surface wave solutions may be found in some of these situations. Potentially, these novel results may be applied in practically useful underwater and underground communication systems in which broadband radio communication channels may be established along the surfaces of rusty pipes deployed in underwater and underground oil and gas producing fields. The developed theory also has important implications in the field of metamaterial optics, where the negative effect of metamaterial losses on signal propagation remains a major problem. The usefulness of the developed approach will be illustrated by several newly found examples of gradient geometries in which the surface wave problem may be solved analytically. And finally, we are going to present several examples of practically useful novel surface wave geometries spanning from radio communication underwater to UV nanophotonics.

## 2 Theory of SEWs in a gradient medium with arbitrary ε(z) and μ(z) dependencies

Let us consider solutions of the macroscopic Maxwell equations in a geometry in which the dielectric permittivity and the magnetic permeability of a medium depend only on z coordinate: ε=ε(z)=ε′(z)+iε″(z) and μ=μ(z)=μ′(z)+iμ″(z), as illustrated in Fig.1. Both ε=ε(z) and μ=μ(z) functions are assumed to be continuous and the effects of spatial dispersion are neglected. These functions are determined by experimental measurements (Note that the typical sources of spatial dispersion in electromagnetism involve anisotropic material response due to crystal structure or optical activity in solutions of chiral molecules. Spatial dispersion is also important in collisional damping in plasmas. None of these factors is present in the experimental situations studied in our work). Under such conditions the spatial variables in the Maxwell equations separate, and without the loss of generality we may assume electromagnetic mode propagation in the x direction, leading to field dependencies proportional to $e^{i(kx-\omega t)}$ (see also ideologically similar derivations in [4,5]). The macroscopic Maxwell equations straightforwardly lead to a wave equation

$$-\nabla^2 \vec{E} - \vec{\nabla}\left(E_z \frac{\partial \varepsilon}{\partial \varepsilon \partial z}\right) = \frac{\omega^2 \varepsilon \mu}{c^2} \vec{E} + (\vec{\nabla}\mu) \times \left(\frac{\vec{\nabla} \times \vec{E}}{\mu}\right) \quad (1)$$

where $E_z$ is the z component of electric field. For the TE polarization this wave equation may be written as

*Corresponding author: Igor I. Smolyaninov, Saltenna, 1751 Pinnacle Drive #600 McLean, VA 22102, USA; igor.smolyaninov@saltenna.com; ORCID 0000-0003-3522-6731
Quirino Balzano, Saltenna, 1751 Pinnacle Drive #600 McLean, VA 22102, USA; quibalzano@gmail.com; ORCID 0000-0002-4384-8402
Vera N. Smolyaninova, Towson University, 8000 York Rd., Towson, MD 21252, USA; vsmolyaninova@towson.edu; ORCID 0000-0003-1959-7698
Daryna Soloviova, Towson University, 8000 York Rd., Towson, MD 21252, USA; dsolov1@students.towson.edu

$$-\nabla^2 E_y = \frac{\omega^2 \varepsilon \mu}{c^2} E_y - \frac{1}{\mu}\left(\frac{\partial \mu}{\partial z}\right)\left(\frac{\partial E_y}{\partial z}\right) \quad (2)$$

If an effective "wave function" $E_y = \psi \mu^{1/2}$ is introduced, the wave equation (2) may be re-written in the form of one-dimensional Schrodinger equation

$$-\frac{\partial^2 \psi}{\partial z^2} + \left(-\frac{\varepsilon(z)\mu(z)\omega^2}{c^2} - \frac{1}{2}\frac{\partial^2 \mu}{\mu \partial z^2} + \frac{3}{4}\frac{(\partial \mu / \partial z)^2}{\mu^2}\right)\psi = -\frac{\partial^2 \psi}{\partial z^2} + V\psi = -k^2\psi \quad (3)$$

On the other hand, for the TM polarization of greater interest to us, the wave equation may be written in the form of a similar one-dimensional Schrodinger equation

$$-\frac{\partial^2 \psi}{\partial z^2} + \left(-\frac{\varepsilon(z)\mu(z)\omega^2}{c^2} - \frac{1}{2}\frac{\partial^2 \varepsilon}{\varepsilon \partial z^2} + \frac{3}{4}\frac{(\partial \varepsilon / \partial z)^2}{\varepsilon^2}\right)\psi = -\frac{\partial^2 \psi}{\partial z^2} + V\psi = -k^2\psi \quad (4)$$

where the effective "wave function" $E_z = \psi/\varepsilon^{1/2}$ has been introduced. In both Eqs. 3 and 4 the $-k^2$ term plays the role of effective energy in the corresponding Schrodinger equation. Note that the continuity of $D_z = \psi \varepsilon^{1/2}$ requires continuity of $\psi(z)$ as long as $\varepsilon(z)$ is continuous and $\varepsilon = \varepsilon' + i\varepsilon'' \neq 0$.

The most interesting result of refs. [4,5] came from the consideration of the TM polarized solutions inside a medium having almost pure imaginary dielectric permittivity $\varepsilon(z) \approx i\varepsilon''(z) = i\sigma(z)/\varepsilon_0 \omega$ where $\varepsilon_0$ is the dielectric permittivity of vacuum, $\varepsilon''$ is very large, and the medium conductivity $\sigma(z)$ is expressed in practical SI units. Based on Eq. 4, the effective potential in such a case may be written as

$$V(z) = -\frac{i\sigma\mu\omega}{\varepsilon_0 c^2} - \frac{1}{2}\frac{\partial^2 \sigma}{\sigma \partial z^2} + \frac{3}{4}\frac{(\partial \sigma / \partial z)^2}{\sigma^2} \quad (5)$$

As was observed in [4,5], the second and the third terms in Eq. 5 are real even in the case of very large $\varepsilon''$, and they may become much larger than the first term if the medium conductivity changes fast enough near the surface as a function of z. If these terms indeed dominate the first one, a propagating SEW solution may appear in a highly lossy system. However, if $\mu$ is real, the first term in Eq.5 (which equals Im($V$)) limits the propagation length of such a SEW solution, unless the frequency is very low.

The key observation of the current work is that the SEW propagation length may improve considerably if $\mu(z)$ also becomes an almost pure imaginary function (while the effective potential well described by Eq.5 remains deep enough to support a SEW mode). This may happen in the case of $\mu$ near zero (MNZ) materials [10], and also in the case of any material in which $\mu' \approx 0$ (while $\mu''$ is not necessarily small). The MNZ conditions are typically observed in split ring resonator (SRR) array metamaterials which effective permeability is defined [11] as

$$\mu_{eff}(\omega) = 1 - \frac{1 - (\omega_s / \omega_p)^2}{1 - (\omega_s / \omega)^2 - i\gamma / \omega} \quad (6)$$

where $\omega_s$ is the resonant frequency of SRR, $\omega_p$ is the magnetic plasma frequency of the array, and $\gamma << \omega_s$ is the damping constant. It is assumed that $\omega_p > \omega_s$ so that $\mu'' > 0$ at all frequencies. The MNZ conditions are observed around $\omega \approx \omega_p$, and in this frequency range

$$\mu_{eff}(\omega) \approx \frac{i\gamma \omega_p}{\omega_p^2 - \omega_s^2} \quad (7)$$

If a gradient medium incorporates such an SRR array metamaterial, the surface potential described by Eq.5 becomes almost purely real:

$$V \approx \frac{\sigma\gamma\omega_p^2}{\varepsilon_0 c^2 (\omega_p^2 - \omega_s^2)} - \frac{1}{2}\frac{\partial^2 \sigma}{\sigma \partial z^2} + \frac{3}{4}\frac{(\partial \sigma / \partial z)^2}{\sigma^2} \quad (8)$$

(note that the practical SI units of conductivity are used in this equation). Moreover, if the damping constant is small, the second and the third terms in Eq. 8 are much larger than the first one, so that the SEW eigenstate is not perturbed.

Beside the SRR metamaterial geometries, the MNZ conditions may also occur in natural materials, such as yttrium iron garnet (YIG) near the ferromagnetic resonance (FMR) [12]. Since FMR is also present in surface oxidized iron [13], our theoretical results may potentially be implemented in practically useful underwater and underground communication systems in which broadband (MHz as compared to KHz range acoustic communication) SEW radio communication channels may be established along the surfaces of rusty pipes deployed in underwater and underground oil and gas producing fields. While this important possibility deserves a separate detailed investigation, let us now concentrate on several analytical SEW problems which may be solved using the outlined general formalism.

## 3 Analytical solutions of SEW problem in gradient media

While in a general arbitrary case the effective Schrodinger equations (Eqs. 3,4) describing the SEW problem in gradient media must be solved numerically (see examples given in [4,5,8]), several important simple geometries may be solved analytically. Let us begin by demonstrating that the gradient terms in Eqs. 3,4 cannot be neglected. Indeed, if we assume that

$$-\frac{1}{2}\frac{\partial^2 \varepsilon}{\varepsilon \partial z^2} + \frac{3}{4}\frac{(\partial \varepsilon / \partial z)^2}{\varepsilon^2} = 0 \quad (9)$$

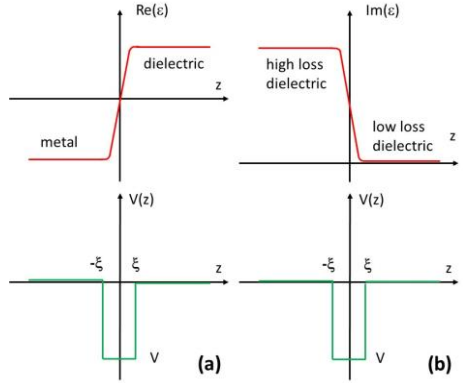

**Fig. 2:** Two kinds of gradient interfaces which produce a rectangular potential well at the interface. (a) A metal-dielectric interface which leads to a surface plasmon-like SEW solution. (b) An interface between low-loss and high-loss dielectric which leads to Zenneck-like SEW solution.

this differential equation may be solved analytically, and the most general solution of this equation is either a constant, or

$$\varepsilon(z) = \frac{c_2}{(c_1+z)^2} \tag{10}$$

where $c_1$ and $c_2$ are constants. It is clear that such a solution cannot approximate any kind of step-like function representing a real interface between two different materials.

Let us now consider two non-magnetic gradient media geometries shown schematically in Fig. 2. Fig. 2a corresponds to a gradual metal-dielectric interface. Let us assume that

$$\varepsilon' = \alpha z \quad \text{for} \quad |z| < \xi \tag{11}$$

and that $\varepsilon'$ stays constant outside this region. We will neglect the second derivative of $\varepsilon'$ near $z=\pm\xi$. We also assume that $\varepsilon''=\delta\gg\varepsilon'$, and it stays constant across the interface. Under these assumptions, in the limit of small $\omega$, the effective potential $V(z)$ given by Eq. 4 becomes a rectangular potential well which depth is:

$$V = -\frac{3\alpha^2}{4\delta^2} \tag{12}$$

This potential well has at least one bound SEW state which corresponds to a surface plasmon-like SEW solution. Note that the usually derived SPP solution [1] corresponds to the limit $\xi\to 0$ and $\delta=0$ in which $\varepsilon(z)$ becomes discontinuous, so that $V\to-\infty$ and the continuity of $\psi$ at $z=0$ also cannot be enforced. In such a special case the derivation of conventional SPP solution reverts to its more usual derivation in [1].

In another example presented in Fig. 2b, a gradient interface between a high-loss and a low-loss dielectric is considered, so that

$$\varepsilon'' = -\alpha z + \alpha\xi \quad \text{for} \quad |z| < \xi \tag{13}$$

and $\varepsilon''$ stays constant outside this region. Once again, we neglect the second derivative of $\varepsilon''$ near $z=\pm\xi$. We also assume that $\varepsilon'=\delta\gg\varepsilon''$, and it stays constant across the interface. Under these assumptions the effective potential $V(z)$ given by Eq. 4 once again becomes a rectangular potential well which depth is also defined by Eq. 12. Similar to the case of Fig. 2a, this potential well also has at least one bound SEW state which corresponds to a Zenneck-like SEW solution.

In principle, almost any 1D surface potential $V(z)$ may be emulated by proper (metamaterial) engineering of either $\varepsilon(z)$ or $\mu(z)$ profile. Let us consider a 1D Coulomb potential and the case of engineered $\varepsilon(z)$ as an example. The needed profile may be obtained by solving a differential equation

$$-\frac{1}{2\varepsilon}\frac{\partial^2\varepsilon}{\partial z^2} + \frac{3}{4}\frac{(\partial\varepsilon/\partial z)^2}{\varepsilon^2} = -\frac{\alpha}{z} \tag{14}$$

where $\alpha$ is an arbitrary positive constant. This equation has been solved using ordinary differential equation solver Wolfram Alpha [14], which gives the following approximate solution:

$$\varepsilon(z) = c_2 \exp\left(\int_1^z \frac{c_1 J_1(2\sqrt{\alpha\zeta}) + \sqrt{\alpha\zeta}\left((c_1-2)J_0(2\sqrt{\alpha\zeta}) - c_1 J_2(2\sqrt{\alpha\zeta})\right)}{\zeta J_1(2\sqrt{\alpha\zeta})(1-c_1)} d\zeta\right) \tag{15}$$

where $c_1$ and $c_2$ are constants, and $J_n(z)$ is the Bessel function of the first kind. Since $1/z \to 0$ in the limit of large $z$, this solution tends to a constant in this limit (see discussion of solutions of Eq. 9 above). On the other hand, $\varepsilon(z)$ diverges near $z=0$. Thus, an interface between two media may be engineered as shown in Fig.

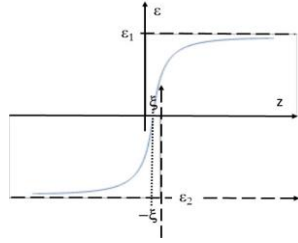

**Fig. 3:** A Coulomb-like potential well at an interface of two media may be constructed approximately from two $\varepsilon(z)$ profiles given by Eq. 15. The $1/z$ divergences are cut off at $z=\pm\xi$.

3, so that near the interface $\varepsilon(z)$ may be approximated with a cutoff 1D Coulomb potential. In the limit of small $\omega$ such a potential well will support a set of Rydberg-like SEW states – see for example ref. [15].

The analytical results outlined above support results of [4,5,8] which indicate that SEW modes are ubiquitous, and they can be found and utilized in all kinds of electromagnetic environments. Below we will illustrate several examples of such applications in the cases of radio communication underwater in the MHz frequency range, Wi-Fi communication through steel pipes in the GHz frequency range, and UV nanophotonics.

## 4 Novel experimental applications of surface electromagnetic waves

### 4.1 Underwater radio communication at MHz frequencies

Let us start with an application of SEWs to underwater radio communication. These experiments were performed using SEW antennas [6] in the MHz frequency range. A SEW antenna excites only the propagation mode supported by a dielectric (or magnetic) discontinuity when it is placed close to it. A SEW antenna does not radiate or support free space electromagnetic fields. In the MHz frequency range underwater they demonstrated efficient radio communication over distances of many hundreds of skin depth. An example of the experimental geometry of a typical 50 MHz underwater radio communication experiment performed in chlorinated pool water (in which skin depth at 50 MHz equals $\eta$=0.1 m) is depicted in Fig.4a. In these experiments two Yaesu VX-8 radios operated at 50 MHz at 5 W output power were connected to their respective SEW antennas (as described in detail in [6]), and used for voice communication between two divers, while both divers and all the components of their radio systems were completely submerged underwater.

Quantitative analysis of these experiments provides strong evidence of the SEW transmission mechanism underwater. When plotted as distance L versus the sum of diver depths $D_1$ and $D_2$ underwater (see Fig. 4b), the experimental datapoints exhibit a linear dependence. The slope of this dependence clearly points at the SEW communication mechanism. Indeed, in the absence of external RF noise, radio communication underwater is limited by the constant radio receiver sensitivity, so that

$$e^{-\frac{D_1}{\eta'}} e^{-\frac{D_2}{\eta'}} e^{-\frac{L}{\eta^*}} \approx const \tag{16}$$

where $\eta'$ is the SEW penetration depth into water, and $\eta^*$ is the SEW propagation length along the water surface (note that neither of these parameters should necessarily be equal to the skin depth η, and in fact both η' and η* exceed η considerably). The logarithm of Eq. 16 gives rise to the linear dependence observed in the experiment:

$$L \approx C - \frac{\eta^*}{\eta'}(D_1 + D_2) \quad (17)$$

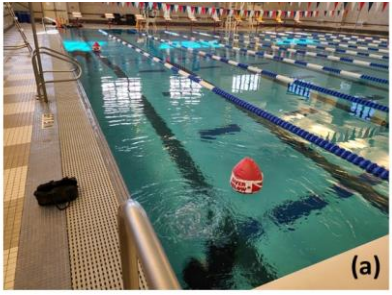

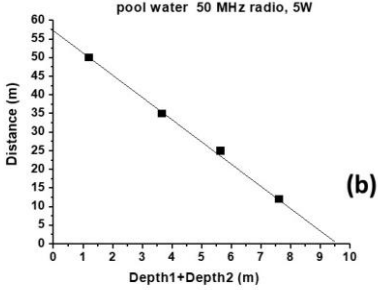

**Fig. 4:** (a). Photo of an underwater radio communication experiment performed at 50 MHz in pool water (skin depth η =0.1m). The lateral positions of two divers underwater are marked by two red plastic buoys. (b) Experimental results plotted as communication distance vs. the sum of depths of the two radios underwater. Note that radio communication was achieved over distances of many hundreds of skin depth.

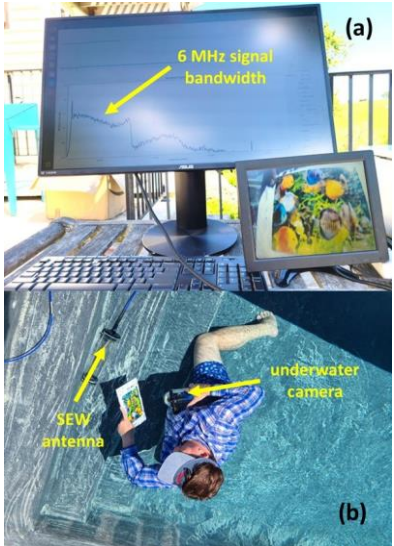

**Fig. 5:** Transmission of live high-definition video image underwater using SEW antennas operating at 30 MHz carrier frequency. As indicated in (a), the video signal bandwidth is 6 MHz.

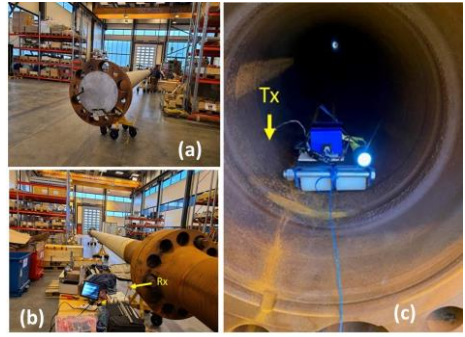

**Fig. 6:** SEW mediated transmission of Wi-Fi video signal through a 32 m long section of a steel riser pipe. The pipe was tightly closed on both ends with thick aluminium shields (a). The Wi-Fi video transmitter was moved through the length of the pipe, while live video (seen on the monitor in image (b)) was transmitted from inside the pipe (c) with no interruption.

Based on the experimental datapoints in Fig.4b, it appears that η*/η'>6, which confirms the SEW-mediated mechanism of the radio communication links underwater, and which clearly explains the observed combinations of radio communication depths and distances far beyond the conventional limit set by the bulk skin depth η of pool water. Note that at 50 MHz the pool water should be considered a highly lossy dielectric. Since pool water conductivity typically measures σ≈0.5 S/m, the imaginary part of its dielectric permittivity ε"≈180 considerably exceeds its real part ε'≈81. The ability of SEW-based underwater radios to reach communication distance of the order of 60 m, or about 600 skin depth (and about 100 wavelengths in pool water at 50 MHz) clearly illustrates low-loss propagation of the SEW modes compared to the bulk radio wave propagation underwater. We should also note that due to its very large bandwidth (6 MHz at 30 MHz carrier frequency), SEWs enable the transmissions of live high-definition video signals underwater (see Fig. 5 and a detailed report in [7]), which is impossible to achieve using more traditional acoustic communications underwater, which are typically operated in the KHz range [16].

### 4.2 Radio communication through metal at GHz frequencies

Another interesting application of the developed theoretical formalism is the case of common metals in the GHz frequency range. All metals behave in this range as very high loss dielectrics, since their dielectric properties are dominated by their very high conductivities in the typical range of $10^6$-$10^7$ S/m. Indeed, recalling the relationship between ε" and material conductivity σ mentioned above:

$$\varepsilon''(z) = \sigma(z)/\varepsilon_0 \omega \,, \quad (18)$$

we see that the case of metals in the GHz range basically corresponds to the case of seawater in the KHz and lower MHz ranges. The easiest way to illustrate the capabilities of GHz range

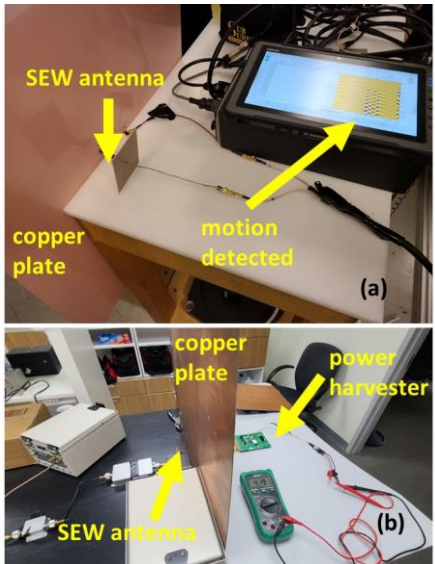

**Fig. 7:** (a) Plasmonic antenna connected to a commercial GPR (QM1020 made by USRadar) detects hand motion behind a copper plate. (b) Plasmonic antenna connected to a commercial power harvester (P2110-EVAL-01 made by Powercast Corporation) enables wireless power transmission through a copper plate.

SEWs in the high-loss metal geometries is to consider propagation of SEW-mediated signals through metal enclosures. SEWs are known to greatly enhance signal transmission through arrays of deeply subwavelength holes made in metal surfaces [17]. The physical mechanism of this effect is believed to involve coupling of incoming electromagnetic waves to SEWs propagating on the opposite metal-dielectric interfaces of a metal layer. While the original observations in ref.[17] were made in the visible frequency range, virtually similar effects were recently observed using SEW antennas operating in the GHz range [18]. Several fascinating examples of practically useful applications of this effect are presented in Figs. 6,7. Fig.6 illustrates transmission of live Wi-Fi video signal through a steel riser pipe of a kind typically used by oil and gas industries, which was tightly closed on both ends with thick aluminium shields. Note that the pipe is visibly rusty, which means that both gradients of $\varepsilon(z)$ and $\mu(z)$ must be taken into account in the complete theoretical description of this experiment. Fig. 7 gives further examples of GHz-range SEW signal transmission through metal plates. In the first one illustrated in Fig. 7a a SEW antenna was connected to a commercial ground penetrating radar (GPR) so that motion detection was enabled through a metal plate. A very clear signal is seen in Fig.7a which was produced by a slight hand wave behind the copper plate. It was verified that the native conventional antennas of this GPR were unable to detect any similar hand motion behind the same copper plate.

As an obvious extension of this experiment, we have verified that wireless power transmission through metal walls is also possible in the "SEW mode" – see Fig. 7b. In this experiment we have used the power harvester circuit P2110-EVAL-01 made by Powercast Corporation. The 2.45 GHz SEW antenna in contact with a large copper shield was used on the receive side of the experiment. On the transmit side we have used the same SEW Wi-Fi transmitter as in Fig.6, which signal was additionally amplified to 10 W by a commercial outdoor Wi-Fi amplifier. The transmit SEW antenna was placed on the other side of the large copper shield in close contact with metal surface. As seen on the multimeter screen in Fig. 7b, the power harvester circuit achieved 4.19 V charging when connected to the SEW antenna. These experiments show the great promise of SEW antennas and signals in wireless communication and power transfer. Unlike the conventional radio systems, which cannot transmit signals behind metal obstacles, such previously prohibitive situations appear the easiest for the SEW-based remote communication and power transfer.

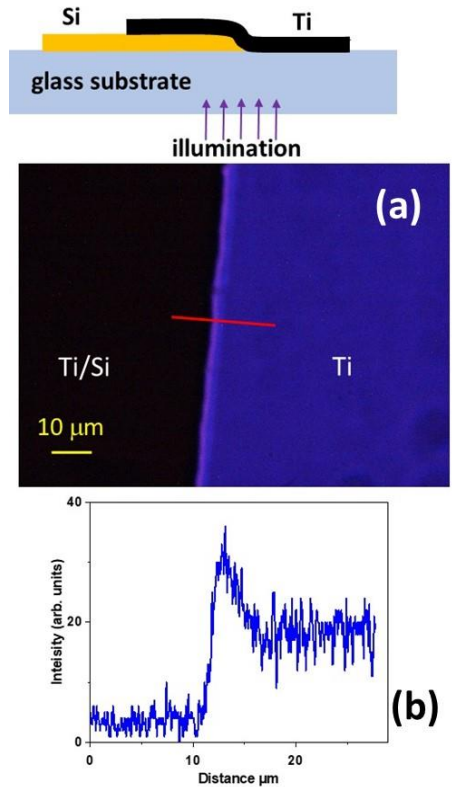

**Fig. 8:** (a) Transmission optical microscope image of an overlap region between a 260 nm thick silicon and a 230 nm thick titanium film. The image size is 117 μm x 88 μm. Note the stripe of enhanced transmission which goes in parallel with the silicon-titanium interface. Its cross section measured along the red line is shown in the plot (b). The top inset in (a) shows experimental geometry.

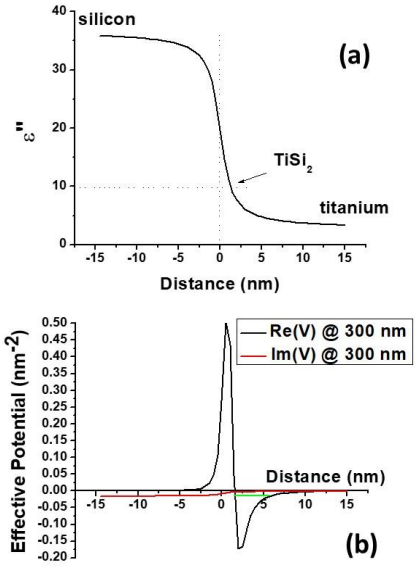

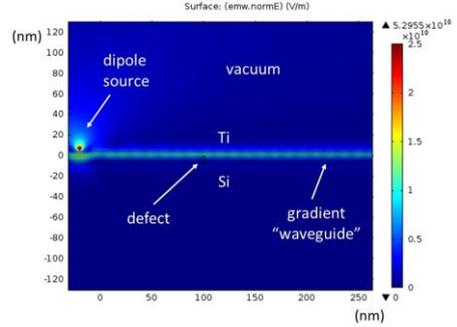

**Fig. 10:** COMSOL Multiphysics simulations of surface wave excitation in a Si/Ti junction at $\lambda_0$=300 nm. The UV light field in the effective gradient waveguide is excited by a dipole radiation source located 7 nm above the junction. A small defect is placed inside the junction at the 100 nm point, resulting in a pattern of standing surface waves. This simulation illustrates SEW propagation over lateral distances which far exceed the SEW wavelength.

**Fig. 9:** (a) Plot of an assumed 20 nm thick planar transition layer between silicon and titanium formed due to diffusion of Ti into Si during deposition. The permittivity of the transition region was assumed to follow a simple arctan law: $\varepsilon''=A\arctan(z/\xi)+B$, where $A=(\varepsilon''_{Si}-\varepsilon''_{Ti})/\pi$ and $B=(\varepsilon''_{Si}+\varepsilon''_{Ti})/2$. The magnitude of $\varepsilon''(z)$ is shown at $\lambda_0$=300 nm. The titanium silicide number is indicated by an arrow. (b) The corresponding effective potential energy (both real and imaginary parts) at the silicon interface defined by Eq. 4 (for TM light) plotted at $\lambda_0$=300 nm. The numerically obtained effective energy level is shown in green.

### 4.3 UV silicon nanophotonics

Finally, let us demonstrate an example of UV nanophotonics applications of the newly discovered SEW solutions in highly lossy dielectric media. Since in the UV range almost all materials behave as highly lossy dielectrics, the TM polarized SEW propagation must exist at a very large number of material interfaces. Such SEW solutions will have a $k$ vector larger than the $k$ vector of photons in each neighbouring medium, thus making them very similar to SPPs. The existence of such large $k$ vector SEW modes in a variety of material geometries may lead to interesting new possibilities in UV nanophotonics and environmental sensing. Unlike the typical plasmonic metals, which are not well suited for nanofabrication using the CMOS techniques, numerous silicon-based SEW geometries will become possible. Indeed, around 300 nm wavelength of UV light, doping of silicon with various metals, such as nickel or titanium, changes the absolute magnitude of the dielectric constant of silicon, while it remains almost pure imaginary [19,20].

We have obtained strong theoretical and experimental evidence in favour of existence of such novel SEW modes in the UV-VIS domain on the example of silicon-titanium interface – see Figs. 8,9. In the experiment illustrated in Fig.8 we have studied UV transmission of an overlap region between a 260 nm thick silicon and a 230 nm thick titanium film. A stripe of enhanced transmission which goes in parallel with the silicon-titanium interface is clearly visible in Fig.8. According to the Fermi's golden rule, the increased transmission of light at the silicon-titanium junction indicates an increased density of electromagnetic states (DOS) in the general area of the junction, which become available to the photons tunnelling through this thick composite conductive film. Therefore, a strong increase in light transmission near the Si/Ti junction is a rather strong indication in favour of the existence of SEW interfacial modes in the UV range.

Numerical modelling of the effective potential $V(z)$ at the silicon-titanium interface (see Fig. 9) performed at 300 nm UV light wavelength also reveals existence of a SEW state. Assuming a 20 nm thick Si/Ti transition region shown in Fig. 9a in which the dielectric permittivity gradually changes from the bulk silicon [19] to the bulk titanium [21] values (compare Fig. 9a with Fig. 3), the resulting effective potential, shown in Fig. 9b, exhibits a pronounced potential well in which $Im(V)<<Re(V)$, and which supports a SEW state indicated in green in Fig. 9b (note a Coulomb-like character of $V(z)$ at $z>0$, which is expected based on Fig. 3). Such a bound state gives rise to at least one solution having almost purely real wave vector $k$, which corresponds to a surface electromagnetic mode with a long propagation length [8].

Fig.10 depicts numerical simulations of surface wave excitation and scattering in a Si/Ti gradient waveguide at $\lambda_0$=300 nm. In these simulations performed using COMSOL Multiphysics solver the UV light field in the Si/Ti gradient surface waveguide is excited by a dipole radiation source placed 7 nm above the junction, and it is scattered by a 4 nm diameter metal defect placed at the 100 nm point, resulting in a pattern of standing surface waves. This simulation illustrates SEW propagation over lateral distances which far exceed the SEW wavelength.

## 5 Discussion and Conclusions

Table 1 below illustrates the common feature of our experiments on SEW propagation in various material systems. In all these situations $\varepsilon''>>\varepsilon'$ in the relevant frequency range, while $\varepsilon''(z)$ exhibits considerable gradients near surface. Note that in the GHz frequency range typical metals are described by complex conductivities, so that $\sigma''\approx\omega\tau\sigma'$, where the inverse relaxation time $\tau^{-1}$ is typically in the infrared frequency range [22]. Therefore, as was noted above, the case of metals in the GHz range basically corresponds to the case of seawater in the KHz and lower MHz ranges. Thus, Table 1 provides us with a strong assurance that three experimental cases considered in Section 4 may be treated in a similar fashion.

**Tab. 1:** Dielectric permittivity of relevant high loss materials.

|     | Pool water@50MHz [6] | Steel@3GHz | Si @300nm [19] |
|-----|---------------------|------------|----------------|
| $\varepsilon'$  | 81                  | ~$10^3$       | 1.7            |
| $\varepsilon''$ | 180                 | 8.7x$10^6$, see (18) | 36       |

Another important observation concerns the $\omega \to 0$ limit of Eqs. 3,4 when they are applied to the case of lossy conductive media, such as water. As demonstrated in [23], in order to make Eq. 18 consistent with Kramers-Kronig relations, in such a case it is necessary for $\varepsilon'$ to possess a delta-function divergence at zero frequency:

$$\varepsilon' = 4\pi^2 \sigma \delta(\omega) + \varepsilon_0 \qquad (19)$$

As a result, while Eqs. 3,4 remain valid at any non-zero frequency, the $\omega \to 0$ limit should be taken with caution. We should note however that the same limitation applies to the standard derivation of the SPP spectra (see for example [1]), since the $\omega \to 0$ limit is also quite complicated even in the case of simple metals [22].

We should also emphasize broader impacts of the developed formalism on the field of PT-symmetric and non-Hermitian optical systems. In particular, it allows fast and straightforward recovery of some important results on PT-symmetric optical waveguides [9]. For example, if spatial dependence of the dielectric permittivity $\varepsilon(z) = c_1 e^{i\alpha z}$ is assumed in Eq. 4, a resulting effective potential $V(z)$ becomes a real constant in the limit $c_1 \omega^2/c^2 \ll \alpha^2$:

$$V(z) = -\frac{c_1 e^{\alpha z} \omega^2}{c^2} + \frac{1}{4}\alpha^2 \approx \frac{1}{4}\alpha^2 \qquad (20)$$

The case of imaginary $\alpha$ in which

$$\varepsilon(z) = c_1 \cos(|\alpha|z) + i c_1 \sin(|\alpha|z) \qquad (21)$$

is illustrated in Fig. 11. It corresponds to a rectangular potential well which supports a symmetric low-loss fundamental mode of such a PT-symmetric waveguide. While making such results quite obvious, we also demonstrated several inherently non-Hermitian electromagnetic systems exhibiting high dissipation, no gain, and no PT-symmetry, which nevertheless have almost real eigenvalue spectrum.

Finally, we should note that neglecting the gradient terms in Eqs. 3,4 will result in the well-studied problem of surface electromagnetic waves supported by a sharp step-like interface [1]. In such a case several low loss SEW solutions are known to be lost [5]. Indeed, the expression for a $k$ vector of the SEW solution in the case of a step-like dielectric interface is well known [1]:

$$k = \frac{\omega}{c}\left(\frac{\varepsilon_1 \varepsilon_2}{\varepsilon_1 + \varepsilon_2}\right)^{1/2} \qquad (22)$$

where $\varepsilon_1$ and $\varepsilon_2$ are the dielectric permittivities of the adjacent media. Eq. 22 may produce an approximately real answer only in a limited number of cases. The first case is a trivial case of both $\varepsilon_1$ and $\varepsilon_2$ being approximately real and positive. The second situation corresponds to surface plasmon polaritons [1], in which case Re($\varepsilon_1$) and Re($\varepsilon_2$) have opposite signs and Re($\varepsilon_1 + \varepsilon_2$)<0. In the third case Im($\varepsilon_1$)$\gg$Re($\varepsilon_2$), which corresponds to the Zenneck wave [2] at an interface between a highly lossy conductive medium and a good low loss dielectric. However, in the latter case, the resulting wave vector appears to be smaller than the wave vector of regular photons in the dielectric, which leads to the "leaky" character of this surface wave.

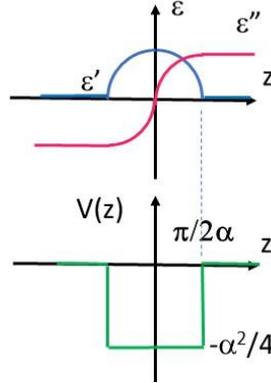

**Fig. 11:** A rectangular potential well formed by a PT-symmetric spatial dependence of $\varepsilon(z) = c_1 e^{i\alpha z}$. Such a potential well supports a symmetric low-loss fundamental mode.

After taking notice of these complications and additional remarks, in conclusion, we have demonstrated that novel low-loss propagating surface wave solutions can exist in a gradient medium in which both dielectric permittivity and magnetic permeability are dominated by their imaginary parts. In addition, we have studied several examples of gradient geometries in which the surface wave problem may be solved analytically. Our results indicate practical ways in which broadband radio communication channels may be established along the surfaces of rusty pipes deployed in underwater and underground oil and gas producing fields. The developed theory also has important implications in the field of metamaterial optics, where the negative effect of metamaterial losses on signal propagation remains a major problem. We also presented several experimental examples of practically useful novel surface wave geometries spanning the range of frequencies from MHz range radio communication underwater to Wi-Fi communication through steel pipes in the GHz frequency range, and to UV nanophotonics.

**Acknowledgments:** The authors are grateful to Mark Barry, Dendy Young, Terence Maxwell, Thomas Snarski, Lukas Hamann, and Jeff Klupt for experimental help.
**Research funding:** N/A
**Author contribution:** All authors have accepted responsibility for the entire content of this manuscript and approved its submission.
**Conflict of interest:** Authors state no conflict of interest.
**Data availability statement:** All data generated or analysed during this study are included in this published article.